\documentclass[aps,superscriptaddress,amsmath,amssymb,floatfix,twocolumn,showpacs,amsfonts,longbibliography]{revtex4-2}
\usepackage{graphicx}% Include figure files
\usepackage{dcolumn}% Align table columns on decimal point
\usepackage{bm}% bold math
\usepackage{physics}
\usepackage{times}
\usepackage[varg]{txfonts}
\usepackage{textcomp}
\usepackage{graphicx}
\usepackage{subfigure}
\usepackage{tabu}
\usepackage{color}
\usepackage[colorlinks=true,citecolor=blue,urlcolor=blue,linkcolor=blue,hyperindex]{hyperref}
\usepackage{braket}
\usepackage{overpic}
\usepackage{amssymb}
\usepackage{multirow}
\usepackage{verbatim}
\title{My Title Here}
\usepackage[]{placeins}
\usepackage{ulem}

\usepackage[usenames,dvipsnames]{xcolor}
\usepackage{tcolorbox}
\usepackage{tabularx}
\usepackage{array}
\usepackage{colortbl}
\tcbuselibrary{skins}
\usepackage{multirow}

\newcolumntype{Y}{>{\raggedleft\arraybackslash}X}

\tcbset{tab1/.style={fonttitle=\bfseries\large,fontupper=\normalsize\sffamily,
colback=yellow!10!white,colframe=red!75!black,colbacktitle=Salmon!40!white,
coltitle=black,center title,freelance,frame code={
\foreach \n in {north east,north west,south east,south west}
{\path [fill=red!75!black] (interior.\n) circle (3mm); };},}}

\tcbset{tab2/.style={enhanced,fonttitle=\bfseries,fontupper=\normalsize\sffamily,
colback=white,colframe=blue!0!black,colbacktitle=Salmon!40!white,
coltitle=black,center title}}

\usepackage{mathtools}
\usepackage{siunitx}
\DeclarePairedDelimiterXPP\BigOSI[2]%
  {\mathcal{O}}{(}{)}{}%
  {\SI{#1}{#2}}

\begin{document}

\preprint{APS/123-QED}
\title{Enhanced mechanical heterogeneity of cell collectives due to temporal fluctuations in cell elasticity} 
%\title{Periodic Stiffening and Softening During Cell Division Increases Mechanical Heterogeneity of 3D Cell Collectives} 
%\title{Temporal dynamics of single cell stiffness determines the invasiveness of an expanding tumor/cellcollective}
\author{Garrett Zills}
\author{Trinanjan Datta}
\email[Corresponding author:]{tdatta@augusta.edu}
\author{Abdul Naseer Malmi-Kakkada}
\email[Corresponding author:]{amalmikakkada@augusta.edu}
\affiliation{Department of Chemistry and Physics, Augusta University, 1120 15$^{th}$ Street, Augusta, Georgia 30912, USA}

%\thanks{A footnote to the article title}%

\date{\today}% It is always \today, today,
             %  but any date may be explicitly specified 
\begin{abstract}
%\mk{hi how r u}
Cells are dynamic systems characterized by temporal variations in biophysical properties such as stiffness and contractility. Recent studies show that the recruitment and release of actin filaments into and out of the cell cortex - a network of proteins underneath the cell membrane - leads to cell stiffening prior to division and softening immediately afterward. In three-dimensional (3D) cell collectives, it is unclear whether the stiffness change during division at the single-cell scale controls the spatial structure and dynamics at the multicellular scale. This is an important question to understand as cell stiffness variations play an important role in tissue spatial organization and cancer progression. Using a minimal 3D model incorporating cell birth, death, and cell-to-cell elastic and adhesive interactions, we investigate the effect of mechanical heterogeneity – variations in individual cell stiffnesses that make up the tumor cell collective – on tumor spatial organization and cell dynamics. We discover that spatial mechanical heterogeneity characterized by a spheroid core composed of stiffer cells and softer cells in the periphery emerge within dense 3D cell collectives, which may be a general feature of multicellular tumor growth. We show that heightened spatial mechanical heterogeneity enhances single-cell dynamics and volumetric tumor growth driven by fluctuations in cell elasticity. Our results could have important implications for understanding how spatiotemporal variations in single-cell stiffness determine tumor growth and spread. 
\end{abstract}

%\pacs{Valid PACS appear here}% PACS, the Physics and Astronomy
                             % Classification Scheme.
%\keywords{Suggested keywords}%Use showkeys class option if keyword
                              %display desired
\maketitle

%\tableofcontents

\section{\label{sec:secI}Introduction}
%Use command \verb!\datta! and \verb!\mk! to insert corresponding comments as
%\verb!\datta{My comment}! to produce \datta{My comment}. 
Changes in cell biophysical properties play fundamental roles in  cancer progression~\cite{mierke2014fundamental,lv2021cell}. Biophysical techniques such as atomic force microscopy (AFM), optical trapping and micropipette aspiration used to probe individual cell mechanical properties~\cite{wu2018comparison} show that cell stiffness grades the ability of tumor cells to metastasize, with softer cancer cells exhibiting the highest migratory and invasive  potential~\cite{swaminathan2011mechanical}. Interestingly, while cancer tissues are generally stiffer than normal tissues, individual cancer cells themselves are softer than normal cells~\cite{cross2007nanomechanical,swaminathan2011mechanical}. 
Given the importance of cell mechanics in cancer progression, insights  %into how temporal variations in the stiffness of individual cells lead 
into how mechanical heterogeneity - i.e. the idea that individual cells within a tumor can be characterized by different stiffnesses - emerge   and consequently impact cell dynamics is important.   
%How time varying single cell stiffness due to  changes in the cell cytoskeletal architecture during cell division determines the mechanical heterogeneity and growth of tumor cell collectives is unclear. 
To address these questions, we used a minimal 3D computational model of cell aggregates 
%encapsulated in a highly viscous medium where cells are modelled as soft spherical agents %As it is of broad interest from both physical and biological perspectives to understand how 
to show that time dependent change in single cell stiffness controls cell dynamics and mechanical heterogeneity of cell collectives. 
%\datta{- The first statement of the third para and this last one is probably taking about related concepts. But, reading this statement and the later, I cannot understand how heterogeneity is related to softeness.}
% https://link.springer.com/article/10.1186/s41236-020-0010-1 

%- Discuss mechanics of cell division and the role of actin cortex \par
Cell division, where a single cell divides into two, is a crucial process in the cell cycle marked by substantial changes in cell morphology, biochemistry and  mechanics~\cite{hurst2021intracellular,nam2018mitotic}. Cell morphological change during division is driven by drastic remodeling of the cytoskeleton - a complex and dynamic network of proteins present in most animal cells%- and stiffening of the cell cortex
~\cite{taubenberger2020mechanics,stewart2011hydrostatic,fischer2016rheology,hurst2021intracellular}. The cell cortex is  composed of a thin actin protein network bound to the cell membrane with a dense crosslinked meshwork architecture~\cite{chugh2018actin} that determines cell deformation in response to intercellular and extracellular forces~\cite{salbreux2012actin}.  
Actin protein filaments that make up the cortex can dynamically polymerize and depolymerize  leading to time dependent variations in  %cell cortex mechanical properties and  consequently 
cell stiffness~\cite{howard2001mechanics,stewart2011force}.  
%- Actin cortex remodeling during cell division can modulate cell stiffness \par 
Recently, %\datta{high - should this word be replaced with something else? e.g. precise, long?} 
high temporal resolution AFM measurements of 
dividing embryonic cells showed that cell stiffness remarkably 
increased immediately prior to cell division and 
softened after cell division, exhibiting a periodic 
stiffening and softening~\cite{fujii2021spatiotemporal}.
% Discuss the time and magnitude of the stiffness change
Cell stiffness increased $\sim$3-fold from $\sim0.1~\mathrm{KiloPascal (KPa)}$ to $\sim0.3~\mathrm{KPa}$ prior to division and softened after division over a time scale of $\sim20$ minutes, which is very short compared to typical cell division times of $15$ hours~\cite{fujii2021spatiotemporal}. %\datta{So, this may be naive question - Does this statement imply that not every periodic softening-hardening sequence is followed by cell division? Maybe this is what was happening, but I never realized it during the calculation process. It may help to clarify this for the reader.} 
Such periodic stiffening and softening is directly driven by the accumulation of actin filaments in the cortical regions immediately prior to cell division and then redistribution into the cytoplasmic regions after division  respectively~\cite{fujii2021spatiotemporal}. A recent study showed that tumor cells exhibit a similar mechanoadaptation by softening to facilitate invasion in confined channels~\cite{rianna2020direct}. 

Using a 3D computational model, we study the effect of rapid single cell level  stiffness change on the overall growth and dynamics of multicellular collectives. 
By varying the probability for cells to soften after division, we discover that mechanical heterogeneity, 3D cell dynamics and tumor growth are all enhanced due to time dependent cell stiffness change.    
%on and softening  probed spatial variation in cell stiffnesses within 3D cell collectives 
%By precisely modeling single cell stiffness change, 
We reveal that cell division associated stiffening and softening determines the spatial structure and dynamics of three-dimensional (3D) multicellular aggregates. Our results provide 
an explanation why softer cells which are directly correlated with heightened cancer progression and metastasis  
are preferentially located at the periphery of multicellular tumor spheroids as observed in  experiments~\cite{swaminathan2011mechanical,lv2021cell,han2020cell}. 
%, how  remains unclear. %Specifically, h is of broad interest from both physical and biological perspectives. 
%\datta{Move this definiton to first para. Relate softening to heterogenity in second para?}. 
%Even though it is of broad interest from both physical and biological perspectives  to understand  
%\datta{usually I provide a very brief summary of the results after asking the main question. Just a suggestion.}
%- Discuss the role of cell stiffness during tumor growth and invasion into the surrounding matrix.
%- Motivate current interest in 3D tumor spheroids and why we turn to modeling in 3D to study cell dynamics
\par %**Mitotic forces~\cite{nam2018mitotic}

\begin{figure*}[hbtp]
\centering
      \includegraphics[width=2.05\columnwidth]{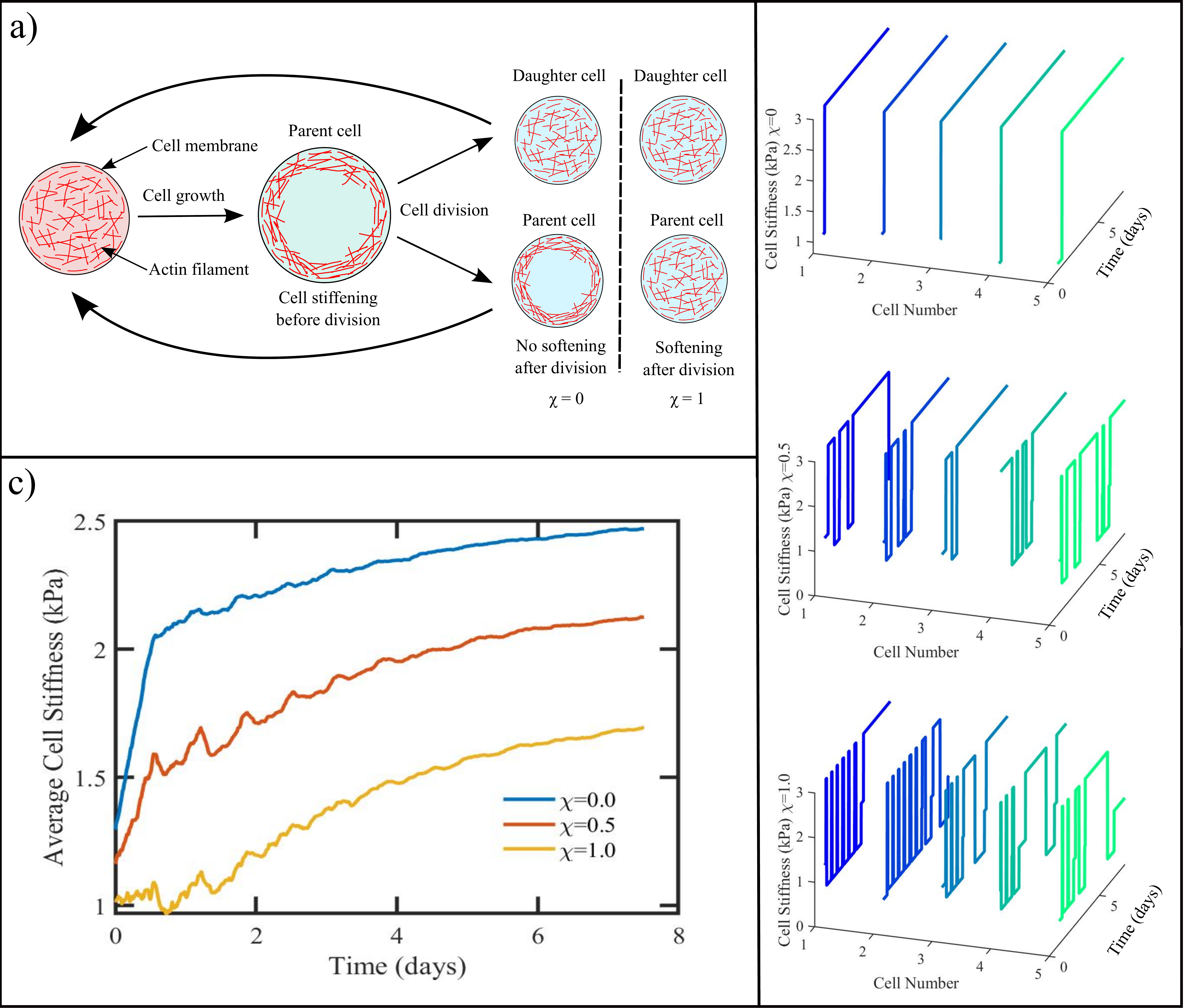}
      \caption{3D tumor growth model with time-varying single cell stiffness.  %\datta{PRB or PRE formatting does not need bold font summary of figure. Summary is okay, just not the font. Also, the figure sublabels can just be (a), ... (b).... no need for bold font} (a) 
      (a) Schematic illustrating time-varying single cell stiffness change implemented in the simulation. The cell stiffness increases prior to division due to the accumulation of actin filaments (red lines) at the cell cortex and soften immediately after division due to the release of cortical actin into the cell cytoplasm, as controlled by the parameter for cell softening probability $\chi$. %\datta{I dont see i and ii in the figure, I also do not see softening only stiffening}. 
      $\chi=0$ implies no softening of the parent cell while $\chi=1$ leads to both parent and daughter cell softening after division. The daughter cell stiffness after division is set from a fixed initial condition. 
      (b) (Top panel) Time dependent individual cell stiffness change at  $\chi=0$. Lines with different colors are for selected individual  cells from the simulation. (Middle panel) Cell stiffness vs time at  $\chi=0.5$ and (Bottom panel) at $\chi=1$. %(c) Time dependent individual cell stiffness change when the probability to decrease in cell stiffness after stiffening is $\chi=0.5$. (d) Time dependent individual cell stiffness change when the probability to decrease in cell stiffness after stiffening is $\chi=1$. 
      (c) Average cell stiffness in the 3D cell collective as a function of time at three different values of $\chi=0,0.5,1$.}%\datta{There are two ds} \datta{The cartoon apsect ratio is off up when combining in PPT. The aspect ratio of figure e) is off. Also, the fonts are inconsistent. I have used TImes New Roman. We just need to be consistent in the axes label, figure label}.
      \label{fig:stiffvstime}
\end{figure*}

This article is organized as follows. In Sec.~\ref{sec:secII} we present the model and the method. In Sec.~\ref{sec:secIII} we characterize the mechanical  heterogeneity of tumor cell collectives and its impact on tumor cell dynamics.  
%In Sec.~\ref{sec:secIV} we present our thermal Hall response results of the multilayer system. 
Finally, in Sec.~\ref{sec:secVI} we present our conclusions. 

\section{\label{sec:secII} Model and Method}

We utilized  
%\datta{wasn't this already developed in your PRX paper? Should we say improved or advanced? There may be a way to say this so that it does not sound like a derivative work} 
an agent based simulation scheme 
for three-dimensional (3D) tumor growth to quantify how time-varying 
single cell stiffness determines the spatial mechanical 
heterogeneity and dynamics of cells within a growing %3D 
multicellular collective.  
%adapted from our previous works on 3D tumor %growth~\cite}. 
Such off-lattice simulations %where individual cells are modeled as soft spherical agents 
are widely used in modelling tumor growth and recapitulate experimentally observed features of %tumor growth and 
individual cell dynamics within cell collectives~\cite{drasdo2005single, schaller2005multicellular, malmi2018cell, malmi2021adhesion, sinha2020spatially, sinha2020self, sinha2021inter}. 
%. In the growing aggregate (see SI, Movies 1-3), 
%whose stiffness changes as a function of time. 
 %The physics of the aggregate growth is governed by two 
%factors - (a) systematic mechanical forces arising from two body interactions, (b) active processes due to particle growth, division and death, as we explain further below. 
%{\bf (b) Active Processes:} 
%Growth of individual particles in the simulation is stochastic %. The growth of the $i^{th}$ particle is 
Agent based models can simulate biophysical interactions between individual cells and provide insight into bridging the gap between single cell and tissue scale behaviors while capturing emergent cell collective behaviors~\cite{schluter2015multi,gorochowski2016agent,malmi2018cell}.  
Cell-cell interactions are typically modelled with short-ranged forces  consisting  of two terms - (i) elastic (repulsion) and (ii) adhesive (attraction) forces.  
The magnitude of the elastic force ($F_{ij}^{el}$) between two cells $i$ and $j$ 
of radii $R_i$ and $R_j$ is given by~\cite{schaller2005multicellular, 
malmi2018cell}, 
\begin{equation}
F_{ij}^{el}=\frac{h_{ij}^{3/2}}{\frac{3}{4}\left(\frac{1-\nu_i^2}{E_i (t)}+\frac{1-\nu_j^2}{E_j(t)}\right)\sqrt{\frac{1}{R_i(t)}+\frac{1}{R_j(t)}}},
\end{equation}
where $\nu_i$ and $E_i$ are the Poisson ratio and elastic modulus of the $i^{th}$ cell. Here, $h_{ij}$ is the overlap (virtual) distance between the two cells.  
The time-varying cell elastic modulus, $E_i(t)$, which we refer to as the cell stiffness is the key parameter that we focus on in this study. Prior works have considered the cell stiffness to be {\it time independent}~\cite{drasdo2005single, schaller2005multicellular, malmi2018cell, sinha2020spatially, sinha2020self, sinha2021inter}.   
The adhesive force $F_{ij}^{ad}$ between cells is, 
\begin{equation}
F_{ij}^{ad}=A_{ij}f^{ad}\frac{1}{2}\left(c^{rec}_i c_j^{lig}+c^{lig}_i c_j^{rec}\right),
\end{equation}
where $A_{ij}$ is the overlap area between the two cells 
in contact and $f^{ad}$ determines the strength of adhesive bond~\cite{schaller2005multicellular, malmi2018cell}. The receptor (rec) and ligand (lig) concentrations are normalized to satisfy $c^{rec}_i=c^{lig}_i=0.9$. 

Starting with 100 cells randomly placed in a 3D cubic volume, we simulate tumor cell 
collective growth over $\sim$7.5 days, sufficient to account for multiple cell division cycles. As cells grow, divide and move the multicellular collective grows into a large spheroid 
with cells in the core and periphery,   
%recapitulating \datta{use a different word than recapitulate?} 
mimicking the growth of tumor spheroids~\cite{sherar1987ultrasound} and  
organoids~\cite{han2020cell} as observed in experiments. 
%The individual 
%cells making up the cell collective grow stochastically in time and 
%undergo division into daughter cells on reaching a critical size.
The effect of forces that cells experience from its 
micro-environment on growth is accounted through the pressure, $p_i$, that 
cells feel due to neighboring cells  
%, making the growth of individual particles micro-environment dependent. 
using the minimal definition~\cite{drasdo2005single, schaller2005multicellular, malmi2018cell}, %of pressure on the $i^{th}$ cell,   
\begin{equation}
    %p_i=\frac{1}{V_{NN(i)}+ V_i}\sum_{j=1}^{NN(i)}{\bf F}_{ij}\cdot d{\bf r}_{ij}. 
    p_i = \sum_{j=1}^{NN(i)} \frac{|F_{ij}|}{A_{ij}}.
\end{equation}
Here, %$V_{NN(i)}= \sum_{j=1}^{NN(i)} \frac{4}{3}\pi R_j^3$ 
the sum is over the nearest-neighbors ($NN$) of the $i^{th}$ cell and $|...|$ denotes the absolute value. 
%. $V_i=\frac{4}{3}\pi R_i^3$ is the volume of the $i^{th}$ particle and $d{\bf r}_{ij}={\bf r}_i-{\bf r}_j $, center to center distance between particles $i$ and $j$. 
If $p_i$ is smaller than a predetermined threshold value, $p_c$, cells grow in size and divide. However, if $p_i > p_c$, the cell becomes dormant which stalls  size growth and  division. A cell can switch between dormancy and growth depending on whether the ratio of $\frac{p_i (t)}{p_c}$ is greater than or less than $1$~\cite{malmi2021adhesion}. %-\datta{one as in number or one as in either or?} respectively. 
The volume of an individual cell grows in time at a mean rate, 
\begin{equation}
    r_V = \frac{2\pi (R_m)^3}{3\tau}, 
\end{equation}
and divides into two cells upon reaching a critical radius $R_m=5~\mu m$.  On division, the parent cell and the newly created daughter cell take on radii  $R_d=\frac{R_m}{2^{\frac{1}{3}}}$ to ensure volume conservation. Hence, a key time scale in the simulation is $\tau$ - the average time it takes for a particle to divide, set to be $\sim15$ hours, comparable to typical cell division times~\cite{casciari1992variations,schaller2005multicellular}.  
%In biological matter such as cells, the cell division cycle is complex process that depends on a network of interacting genes and proteins. In our minimal model, we assume that particle division depends on the critical radius, average division time and pressure. 
We incorporate cell death in the simulations by randomly removing particles at a rate $k_d=10^{-6} s^{-1}$. Owing to the 
death rate being much smaller than the birth rate $(k_d << \frac{1}{\tau})$, we are simulating a rapidly expanding  collection of cells. %In this limit, %here the birth rate ($k_b = 1/\tau$) is larger than the death rate $k_b/k_d \sim 20$ justifies our assumption of a constant death rate. 
%we anticipate our assumption of uniform death rate with no spatial dependence to be feasible because the division rate is much  
%larger than the death rate ($k_b/k_d \approx 20$), where $k_b = 1/\tau$ is the birth rate. 
%As we consider soft spherical particles that allows for partial overlap, we can define an effective shape based on Voronoi tesellation that would be aspherical in the limit where particles are densely packed spatially (see Refs.~\cite{schaller2005multicellular} and~\cite{sinha2020self} for a more detailed discussion).

\subsection{Time variation in single cell stiffness}%\datta{Use subsections in PRB PRE format not bold font sections. This is more for nature journals} 
To investigate whether 
temporal variation in single cell stiffness affects 3D cell collective spatial organization and dynamics,  % we study the impact of time dependent individual cell stiffness during the cell growth cycle.  
we coupled cell division to cell stiffness change %at the single cell level 
according to two simple rules: (i) %mmediately before cell division increasing the cell stiffness prior to division and decreasing cell 
%stiffness immediately post division. 
First, as the size of a cell approaches the mitotic radius, at $R_i(t=t^*)/R_m = 0.98$, its stiffness is increased to 
$E_i(t>t^*)=\mathrm{min}\{2.5\times E_i(t^*),3$KPa\} i.e. minimum of the 
value between $2.5$ times the cell stiffness at time $t^*$ %when $R_i/R_m = 0.98$ 
and a threshold stiffness value of 3 KPa.  %\datta{experimental or theoretical justification, any reference?}. 
This ensures that the maximum 
cell stiffness is 3KPa and prevents it from increasing to unphysical values. Tumor  cells tend to be stiffer than embryonic cells and the stiffness range we consider have been experimentally measured~\cite{rianna2020direct}. The condition $R_i(t=t^*) = 0.98 \times R_m$ is set to ensure that stiffness change occur immediately before cell division at $R=R_m$. (ii) Second, to mimic experimentally observed cell softening after 
division, we implement a probabilistic protocol for cell softening: 
\begin{itemize}
    \item draw a uniformly distributed random number, $u$,  in the interval (0,1) 
    \item if $u$ is less than or equal to the softening probability parameter $\chi$,  %\datta{who decides this? Should we state that this is an input to the model}, 
    reduce cell stiffness to $E_i(t>t^{'})=\mathrm{max}\{0.2\times E_i(t^{'}),0.5$~KPa\}. If $u > \chi$, the parent cell does not soften. $\chi$ is an input parameter 
    that we vary in the model.     
\end{itemize}
To prevent cell stiffness from approaching zero, we implement a lower bound of cell stiffness at $0.5$~KPa. The initial condition for daughter cell mean stiffness is set to $1$KPa and characterized by a Gaussian distribution with standard deviation of $0.1$ KPa (see Table I).  %in the simulation cell stiffnesses can dynamically vary between $0.5-3$KPa. 
The spectrum of cell stiffness between $0.5-3$KPa we consider is in the physiological range for cell  stiffnesses~\cite{plodinec2012nanomechanical} with the lower and upper end corresponding to embryo cell stiffness and lung cell stiffness respectively~\cite{irianto2016snapshot}. The schematic of single cell stiffness change is visualized in Fig.~\ref{fig:stiffvstime}a. The time dependent single cell stiffness change obtained in the simulation for selected cells at three different $\chi$ values are shown in Fig.~\ref{fig:stiffvstime}b. At $\chi=1$ 
intermittent cell stiffening and softening events are visible (bottom  panel, Fig.~\ref{fig:stiffvstime}b) compared to $\chi=0$  (top panel, Fig.~\ref{fig:stiffvstime}b).  
%\datta{experimental or theoretical justification, any reference?}. 

We now describe the molecular underpinnings of the cell stiffness change implemented in the computational model. As cells progress through the cell cycle 
and approach division, actin filaments 
accumulate at the boundary of the cell, increasing cell stiffness. %\datta{I would include the cartoon picture here separately, just a suggestion}.  
After division, the acto-myosin filaments are distributed into the cytoplasmic regions of the cell until the cell is ready for the next 
division event~\cite{fujii2021spatiotemporal,ramanathan2015cdk1,kelkar2020mechanics}. The probability for the cortical protein filaments to be redistributed into the cytoplasm is modeled in our simulation via the parameter $\chi$, which we vary from $0$ to $1$, at intervals of $0.1$. Hence, $\chi=0.1$ implies a very low probability for cells to soften, while $\chi=1$ implies a high probability for cell softening after division. We note that the parent cell stiffness is dynamically increased before division. After division, the daughter cell stiffness is set from the initial condition as noted in Table I while the parent cell undergoes softening as determined by the parameter $\chi$.   
When the daughter cell grows in size and approaches the mitotic radius as noted above ($R_i(t=t^*)/R_m = 0.98$), they can undergo stiffening followed by softening.  Furthermore, some cells may become dormant after stiffening and not progress to division depending on the pressure parameter. This would lead to the arrest of single cell stiffnesses at heightened values irrespective of $\chi$. The overall change in average cell stiffness of the cell collective as a function of time is 
shown in Fig.~\ref{fig:stiffvstime}c. At $\chi=1$, the cell collective is on 
average the softer as opposed to a stiffer cell collective at $\chi=0$. Based on 
varying the parameter $\chi$ we can now study the impact of cell division associated cell softening and stiffening on cell dynamics within the 3D cell collective. 
% \datta{- dont understand this statement}. 
%\textbf{Table I:}
%The parameters used in the simulation. 
\par
\begin{table}
\begin{tabular}{ |p{4.5cm}||p{3cm}|p{1cm}|  }
 \hline
 \bf{Parameters} & \bf{Values}  \\
 \hline
 Timestep ($\Delta t$)& 10$\mathrm{s}$  \\
 \hline
 Critical Radius for Division ($R_{m}$) &  5 $\mathrm{\mu m}$ \\
 \hline
 Environment Viscosity ($\eta$) & 0.005 $\mathrm{kg/ (\mu m~s)}$  \\
 \hline
 Average Division Time ($\tau$)  & 54000 $\mathrm{s}$\\
 \hline
 Adhesive Coefficient ($f^{ad})$&  $1\times 10^{-4} \mathrm{\mu N/\mu m^{2}}$\\
 \hline
 Initial Mean Elastic Modulus ($E_{i}$) (Standard Deviation) & $1 \mathrm{KPa} (0.1 \mathrm{KPa})$ \\
 \hline
 Mean Poisson Ratio ($\nu_{i}$) (Standard Deviation) & 0.5 (0.02) \\
 \hline
 Death Rate ($k_d$) & $10^{-6} \mathrm{s^{-1}}$ \\
 \hline
 Mean Receptor Concentration ($c^{rec}$) (Standard Deviation)  & 0.9 (0.02) [Normalized] \\
\hline
 Mean Ligand Concentration ($c^{lig}$) (Standard Deviation) & 0.9 (0.02) [Normalized]\\
\hline
 Threshold Pressure ($p_c$) & $10^{-1} \mathrm{KPa}$ \\
\hline
\end{tabular}
\label{table_1}
\caption{The parameters used in the simulation. The parameters where we indicate the mean and standard deviation 
are sampled from a normal distribution. For details, see~\cite{malmi2018cell}.}
\end{table}

\subsection{Cell dynamics} 
In addition to the active forces due to cell growth and division, the 
passive forces experienced by a cell due to interaction with its 
neighbors contributes to cell dynamics. 
The net force ${\bf F}_i$ on the $i^{th}$ cell is the 
vectorial sum of elastic and adhesive forces that the neighboring 
cells exert on it, ${\bf F}_i=\sum_{j=1}^{NN(i)}{\bf F}_{ij}$. 
Here, $j$ is summed over the number of nearest neighbors $NN(i)$. We 
performed over damped (low Reynolds number \cite{purcell1977life}) 
dynamics without thermal noise because the viscosity of the medium surrounding cells is assumed to be 
large~\cite{malmi2018cell}. Therefore, the equation of motion for %the position of the $i^{th}$ 
a cell is,  
 \begin{equation}
\dot{\textbf{r}}_i=\frac{{\bf F}_i}{\gamma_i},
\end{equation}
where $\gamma_i=6\pi\eta R_i$ is the friction term which models the environment as a thick gel. At least $12$ simulations each for $10$ different values of the 
cell softening parameter $\chi$ were conducted %varying from $0$ to $1$ and 
to study its impact on 
single cell dynamics, size of the cell collective and spatial mechanical 
heterogeneity. The various parameters utilized in the computational model are summarized in Table I. 

There are two timescales that are important in our computational model (i) elastic cell-cell interaction time scale $\tau_{el}\sim\frac{\gamma}{ER}$ of $\BigOSI{100}{sec}$ and (ii) the time scale $\tau$ associated with the cell division of $\BigOSI{50000}{sec}$. As we are interested in the long-time dynamics of the cell collective at time scales greater than $\tau$, we consider the time taken for the cell to stiffen and soften to occur instantaneously given that it is a fast  process compared to cell division. We note that the growing spheroid that we consider is a non-equilibrium system~\cite{malmi2018cell,sinha2020spatially,sinha2022mechanical}.  
%,  justifying our assumption that cell stiffening and softening occurs instantaneously \datta{I don't understand why neq is instantaneous?}. 
%~\ref{table_1}.  %\datta{Table I or II?}. 
%However, in this study we assume particle shape to be spherical in order to keep the problem tractable while %for the purposes of calculating forces while 
%in living active matter cell shape can be a highly complex and dynamic quantity~\cite{charras2008life, keren2008mechanism, bi2016motility}. 
%} %mention in the reply regarding voronoi volume %and ${\bf r}_i$ is the position of the $i^{th}$ cell.
 
\section{\label{sec:secIII}Results}
\begin{figure*}[hbtp]
\centering
      \includegraphics[width=2.0\columnwidth]{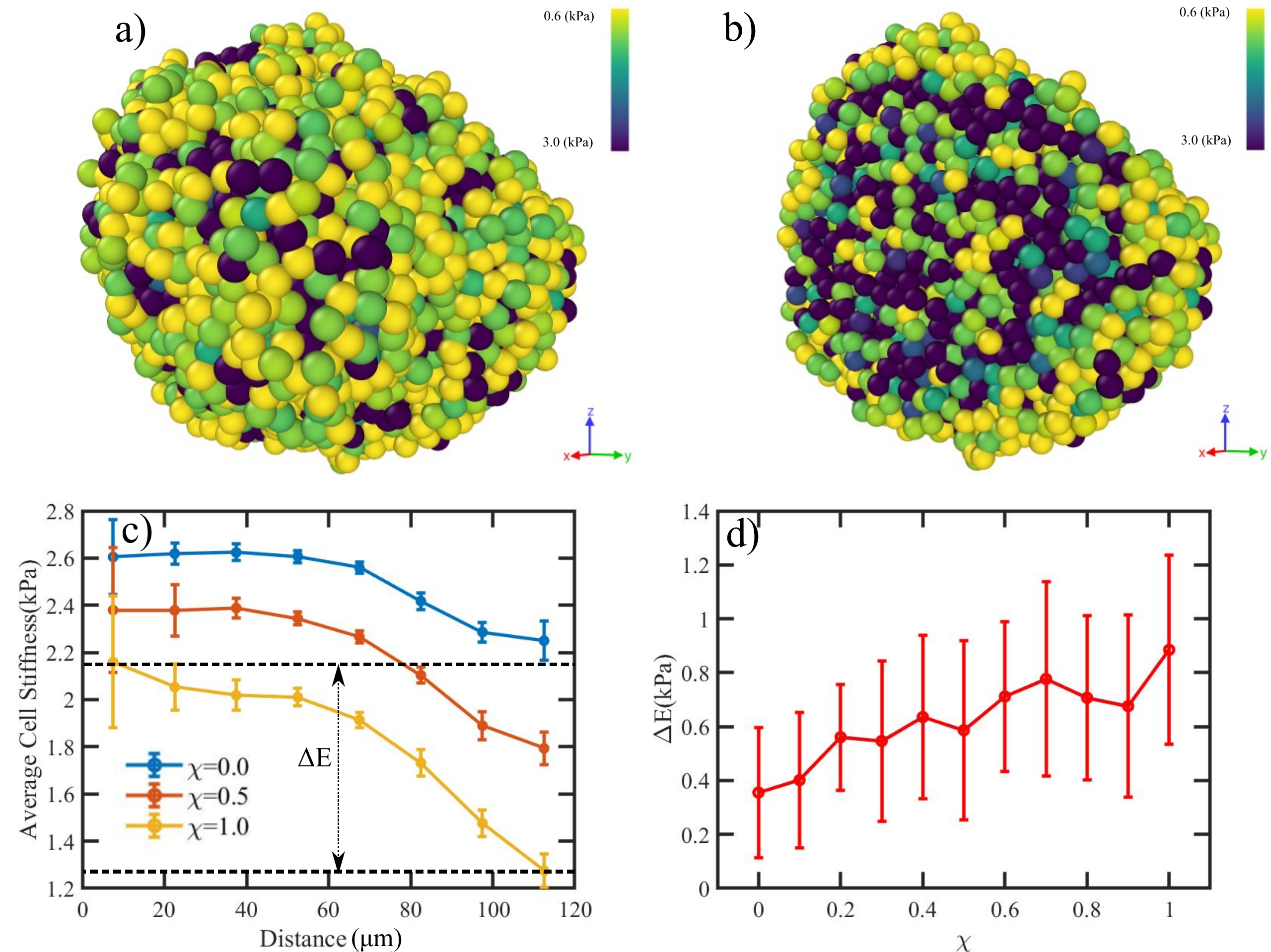}
      \caption{Spatial heterogeneity in cell subpopulation stiffness between core and periphery in growing 3D cell collectives. (a) Snapshot of the 3D collection of $\sim6,000$ cells at $t=7.5$ days for $\chi=1$. Each small sphere is a single cell of maximum diameter $10~\mu m$ with the color visualizing cell stiffness (see color bar). (b) Cross section through one plane of the 3D cell collective showing the core and periphery. Stiffer cells (darker color) are visible at the core with softer cells at the periphery. 
      (c) Average stiffness of cell subpopulations as a function of distance from the core. Cell subpopulations are categorized according to their  distances from the center of mass of the 3D cell collective. Circles indicate mean values and the error bar is the standard deviation. A marked difference between cell subpopulation stiffness at the core vs periphery is noted at $\chi=1$, as quantified by $\Delta E$. (d)  Mechanical heterogeneity of the cell subpopulation stiffness is quantified using $\Delta E$.  Difference in the average cell stiffness between the core and periphery is most pronounced at $\chi=1$. %~\datta{spacing between axis label and units needed}
      }
      \label{fig:mechheter}
\end{figure*}

%\datta{The results section writing needs to be broken up into more paragraphs.}
%We start the simulation with a group of 100 randomly placed cells within a highly viscous 3D environment. We 
%track individual cell positions and stiffnesses as the multicellular 
%cluster grows over approximately 8 days to over $\sim10,000$ cells. 
% \begin{figure*}[hbtp]
% \centering
%       \includegraphics[width=15.8cm]{fig2.pdf}
%       \caption{{\bf Spatial heterogeneity in cell subpopulation stiffness between core and periphery in growing 3D cell collectives.} {\bf a}, Average stiffness of cell subpopulations as a function of distance from the core. Cell subpopulations are categorized by their distances from the center of mass of the 3D cell collective. Circles indicate mean values and the error bar is the standard deviation. A marked difference between cell subpopulation stiffness at the core vs periphery is noted at $\chi=1$. {\bf b}, Mechanical heterogeneity of the cell subpopulation stiffness between the core and periphery is quantified using $\Delta E$ which is the difference between the average cell stiffness at the core and periphery.}
%       \label{fig:mechheter1}
% \end{figure*}
While the molecular factors that determine tumor growth is better understood, much remains to be known about the impact of time dependent 
changes in cell physical properties on the 
spatial 
mechanical heterogeneity of 3D cell collectives. 
Given that individual cells that make up a tumor can be characterized by broadly varying  stiffnesses, are cell subpopulations i.e. clusters of cells with differing levels of stiffness spatially organized within cell collectives? 
To answer this question we visualized the multicellular spheroids generated from our simulations at $t=12 \tau$ and  $\chi=1$ (Fig.~\ref{fig:mechheter}a). A mixture of soft (lighter color) and stiff cells (darker color)  
%with more of the softer cells 
are visible on the surface of the spheroid. %as compared to the spheroid generated at $\chi=0.1$. 
As we are interested in  understanding the spatial variation in cell stiffness,  
%between the core and periphery of the cell collective, 
a cross-section view with respect to a 2D plane cutting through the center of the 3D cell collective is shown in Fig.~\ref{fig:mechheter}b. 
%Figs. at $\chi=1$ and $\chi=0.1$ respectively. 
Remarkably, a clear trend in 
spatial mechanical heterogeneity with stiffer cells in the core and softer cells at the periphery is visible. 
%$\chi=1$. 
%Comparatively, when the probability to soften after division is  supressed at $\chi=0.1$, such a clear stiffness gradient between the core and periphery is not visible (see Fig.). %add a figure to appendix? 
To quantify this further,     
%spatial mechanical heterogeneity of the 3D cell collective, 
we grouped cells 
according to their positions with respect to the tumor center of mass, $R_{CM} = \frac{1}{N}\Sigma_i^N r_{i}$, 
where $N$ is the total number of cells. 
By calculating the cell distances from the tumor center of mass, 
$d_i=|\vec{r}_{i}-\vec{R}_{CM}|$, where $|...|$ indicates vector magnitude we group cells into 8 cell 
subpopulations. Cells closest to the center of mass compose the core of the spheroid and we refer to the outermost subpopulation as the periphery. 
The thickness of each layer composing the cell subpopulation is set to 15 $\mu m$. The statistical average of single cell stiffness within each subpopulation is computed at time $t=12 \tau$ using, 
\begin{equation}
    \langle E(r_d) \rangle = \frac{\Sigma_i E_i \delta(r_d-d_i)}{\Sigma_i \delta(r_d-d_i)}, 
\end{equation}
where $r_d$ is the binning distance from the tumor center of width 15 $\mu m$. %\datta{is center the correct word?}. 

Notably, cells located near the core of the tumor spheroid are  stiffer as compared to cells near the periphery 
(Fig.~\ref{fig:mechheter}c), irrespective of the value of 
%probability to decrease in stiffness after cell division(
$\chi$. As expected, the overall cell subpopulation 
is stiffer at low $\chi$ which corresponds to low probability for cells to  soften post division. We discover that  the stiffness heterogeneity between cell subpopulations in the core and periphery increases with $\chi$. 
%the probability for cell stiffness to decrease after division. 
To quantify the spatial mechanical heterogeneity between cell subpopulations, 
we calculated the difference in average stiffness between the core and 
periphery, 
\begin{equation}
\Delta E=\langle E \rangle_{\mathrm{core}}-\langle E 
\rangle_{\mathrm{periphery}} 
\end{equation}
(see Fig.~\ref{fig:mechheter}d). At $\chi=0$, the 
spatial mechanical heterogeneity is low with a mean $\Delta E \sim0.35$KPa as compared to $\Delta E \sim0.8$KPa at $\chi=1$. 
The spatial mechanical heterogeneity is therefore enhanced at  $\chi=1$, indicating that time-varying cell stiffness change during cell division 
is an important determinant of mechanical heterogeneity. 
Indeed, mechanical heterogeneity during disease progression is thought to facilitate metastasis~\cite{plodinec2012nanomechanical,shen2020detecting,lv2021cell}. Moreover, spatial heterogeneity of tumor organoids with a stiffer core and softer periphery of cells may be a general feature of 3D tumor cell  collectives~\cite{han2020cell}. 
%, in agreement with our simulation results 

\begin{figure*}[hbtp]
\centering
      \includegraphics[width=2.05\columnwidth]{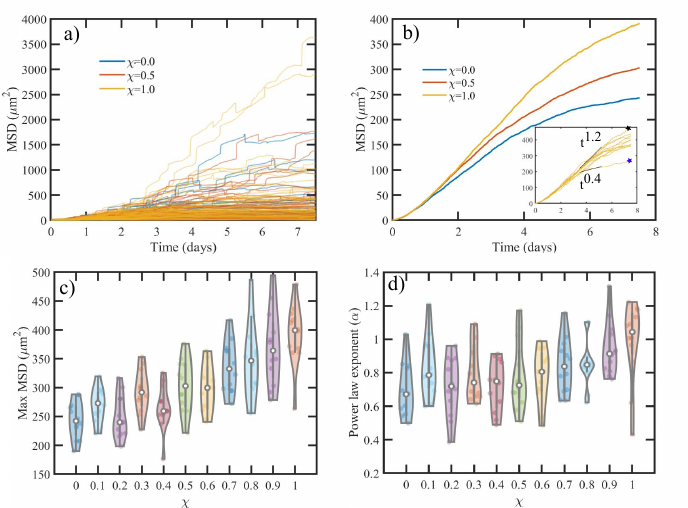}
      \caption{Cell softening after division leads to distinct cell dynamic behaviors. (a) Single cell MSD (scMSD) versus time, at $\chi=0$ (blue), $\chi=0.5$ (red) and $\chi=1$ (yellow). $\sim$ $60$ scMSDs per $\chi$ value show highly heterogenous dynamics with some cells traversing large distances while other cells move less in comparison to the typical cell diameter of $10\mu m$. (b) Ensemble averaged MSD of cells versus time at three different values of $\chi$. 
      The data is averaged over 12 independent simulation runs by tracking $\sim 800$ cells over the complete simulation at each value of $\chi$. Inset: MSD from averaging over cells from individual simulation runs at $\chi=1$. The maximum (max) MSD value for 2 individual  simulation run is marked with stars. The time regime where MSD is fit to power law in order to extract the MSD exponent is shown. (c) Max MSD increases as a function of $\chi$. Colored dots represent each of the Max MSD values from individual simulation runs. White dots are the median values and the thick line within the violin distribution represent the interquartile range between the first and third quartiles. The bottom and top edge of thinner gray 
      lines mark the lower and upper adjacent values respectively. 
      (d) By fitting cell averaged MSD in each of the 12 simulation runs to a power law, we extracted the MSD exponent ($\alpha$) as a function of $\chi$. Cell dynamics is significantly enhanced at $\chi=1$ as compared to $\chi=0$.}
      \label{fig:msd}
\end{figure*}

Next, we investigated whether the dynamics of individual cells that make up the spheroid could be affected by the spatial 
mechanical heterogeneity. %in the cell subpopulation stiffnesses. 
Prior studies report that metastatic tumor cells are softer compared to non-metastatic tumor cells~\cite{guck2005optical,xu2012cell,fritsch2010biomechanical,alibert2017cancer,han2020cell}. As cell division events fluidize 
cell collectives and lead to superdiffusive 
dynamics~\cite{malmi2018cell,sinha2020spatially}, division dependent cell softening could affect the non-equilibrium active forces that cells experience and thus affect individual 3D cell dynamics. By tracking single cell  trajectories, we calculate both single cell mean-squared displacement (scMSD) and ensemble averaged MSD, 
\begin{equation}
\Delta (t) = \left\langle \frac{1}{N} \Sigma_{i=1}^{N} [{\bf r}_{i}(t)-{\bf r}_{i}(0)]^2 \right\rangle,
\end{equation}
where $N$ denotes the total number of tracked cells from the beginning to the end of the simulation. The ensemble average $\langle ... \rangle$ is over 12 different simulation runs at each value of $\chi$ for different initial conditions (see Appendix~\ref{sec:appendixA}).  
scMSD (without averaging over $N$ or multiple simulation runs) shown in Fig.~\ref{fig:msd}a reveal that distances traversed by cells are highly heterogenous. While majority of the cells traverse distances less than $\sqrt{500\mu m^2}=22.4\mu m$, a population of highly dynamic cells exist that traverse distances on the order of $\sqrt{3000 \mu m^2}=54.7 \mu m$ (Fig.~\ref{fig:msd}a) at $\chi=1$. Another interesting feature is the intermittent change in scMSD clearly visible in the highly dynamic group of cells where there are steep increases in scMSD followed by time regimes where scMSD does not change much. 

As  cell collectives exhibit glass to fluid-like transition due to cell division events~\cite{Ranft10PNAS,matoz2017cell,malmi2018cell}, %-when did we figure this out in our project?}
\begin{figure}[ht]
\centering
\includegraphics[width=9cm]{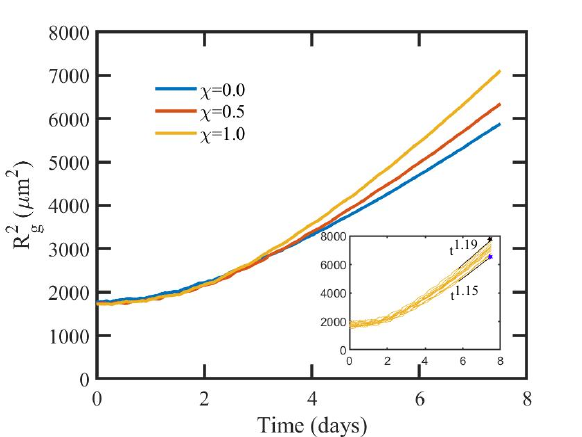}
\caption{Cell softening after division control cell collective
growth. (a) Quantification of the ensemble averaged (over 
12 simulation runs) radius of gyration 
squared ($R_g^2$) of the 3D tumor cell collective over 7.5 days at 
$\chi=0,0.5 ~\mathrm{and} ~1$. Inset: $R_g^2$ from averaging over cells from individual simulation runs at $\chi=1$. Maximum (max) $R_g^2$ values  for 2 individual simulation runs are marked with stars. The time regime where $R_g^2$ is fit to power law in order to extract the exponent is shown. 
% (b) Max value of $R_g^2$ at the end of 
% simulation run at $t=7.5$days indicate significantly enhanced growth 
% of tumor cell collective with increased probability of individual 
% cells to soften (compare $\chi=1$ to $\chi=0$). Max $R_g^2$ 
% value from each of the 12 simulations after averaging over individual 
% cell $R_g^2$ is shown as colored dots. White dots are  the median values and the thick line within the violin distribution represent the interquartile range between the first and third quartiles. The bottom and top edge of thinner gray lines mark the lower and upper adjacent values respectively. (c) By fitting the average $R_g^2$ in each of the 12 
% simulation runs, we extracted the $R_g^2$ exponent $\beta$ as a function of  $\chi$. Volumetric growth of the 3D cell collective is significantly enhanced at $\chi=1$ as compared to $\chi=0$.
}%\datta{The images combined in PPT have different scales. They all need to be the same size. When generating images in MATLAB the figure size can be controlled, I think}}
\label{fig:rg2a}
\end{figure}
we surmise that %the %\datta{net, individual, collective?} 
cell displacement could be linked to  
cell division events and the associated time dependent change in stiffness. Hence, we  investigate the effect of cell softening probability on ensemble averaged MSD (Fig.~\ref{fig:msd}b). At short times, $t<2$ days, probability of cell softening ($\chi$) has no visible effect on the cell dynamics as observed from the MSD plots. %(Fig.~\ref{fig:msd}b). 
In contrast, at longer times $t>2$ days, cell dynamics is significantly restricted %in the 
at low $\chi$. %implying that 3D cell dynamics the spatial mechanical heterogeneity decreases. 
As the MSD is significantly enhanced at $\chi=1$, %when the spatial mechanical heterogeneity is also higher 
we confirm that higher $\chi$ values resulting in  enhanced spatial mechanical heterogeneity with larger stiffness asymmetry between cells in the core and the periphery (see Fig.~\ref{fig:mechheter}d) is more conducive to 3D cell dynamics.   
%stiffer cell collective at $\chi=0$ as 
%compared to tumor cell collectives composed of softer cells at $\chi=1$. 
To confirm that the space explored by cells increases with $\chi$, we analyze the maximum (max) MSD at $t=12\tau$ %\datta{this is 7.5 days which is greater than a week, is that okay, or do we need to be careful previously when we say a week} 
during each individual simulation run (see stars in the inset of Fig.~\ref{fig:msd}b). The max MSD at multiple $\chi$ values are summarized in Fig.~\ref{fig:msd}c.  
Therefore, on the basis of the spatial mechanical heterogeneity we report in Fig.~\ref{fig:mechheter}, a stiffer core and softer peripheral cells is conducive to heightened cell dynamics as indicated by the larger MSD values. 

%mechanical heterogeneity is correlated with increased 
%MSD of particles  depend not only on the active driving forces, but also 
MSD depends on the mechanical resistance of the surrounding medium~\cite{brangwynne2009intracellular}, but, the influence of individual particle level change in mechanical properties such as stiffness %\datta{not sure what the word mechanics is referring to here, motion, properties etc} 
on MSD is unclear. Time dependent scaling of MSD based on a fit to power law $\Delta(t)\sim t^{\alpha}$ reveals important features of cell dynamics (see black lines in the inset of Fig.~\ref{fig:msd}b for details). When $\alpha=1$, cells exhibit diffusive random walk. For cells undergoing directed motion, the power law exponent is greater than one ($\alpha>1$) in contrast to restricted cell motion which leads to a sublinear rise in MSD with $\alpha<1$. Interestingly,  median $\alpha$ values show that cells exhibit subdiffusive motion due to time varying stiffness 
change except at $\chi=1$. The median MSD exponent (white circles in 
Fig.~\ref{fig:msd}d) are all below $1$, except at $\chi=1$. Additionally, heightened mechanical heterogeneity leads to enhanced super-diffusive dynamics as  there is a marked increase in median MSD exponent at $\chi=1$ ($\alpha > 1$). For  $\chi < 0.5$, no clear trend in MSD exponent is visible in Fig.~\ref{fig:msd}d even though the max MSD increases in the same range. Despite the fact that all the MSD exponents are characterized by a wide scatter, we observe that enhanced spatial mechanical heterogeneity led to heightened MSD and MSD exponent. 

  \section{\label{sec:secIV}Individual cell softening regulates cell collective growth rate} 
\begin{figure}[ht]
\centering
      \includegraphics[width=8.6cm]{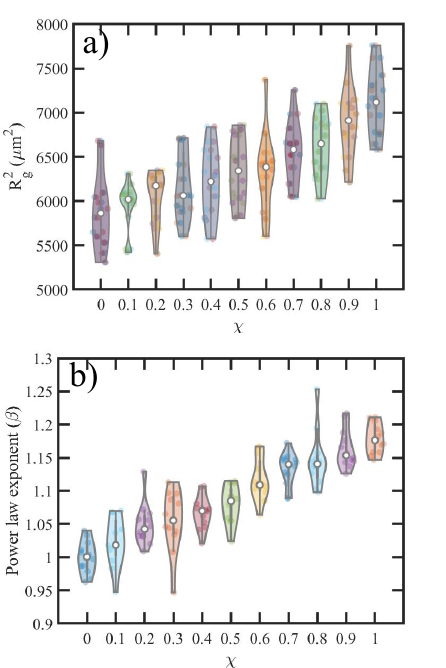}
\caption{Cell softening after division control cell collective
growth. (a) Max value of $R_g^2$ at the end of 
simulation run at $t=7.5$ days indicate significantly enhanced growth 
of tumor cell collective with increased probability of individual 
cells to soften (compare $\chi=1$ to $\chi=0$). Max $R_g^2$ 
value from each of the 12 simulations after averaging over individual 
cell $R_g^2$ is shown as colored dots. White dots are  the median values and the thick line within the violin distribution represent the interquartile range between the first and third quartiles. The bottom and top edge of thinner gray lines mark the lower and upper adjacent values respectively. (b) By fitting the average $R_g^2$ in each of the 12 
simulation runs, we extracted the $R_g^2$ exponent $\beta$ as a function of  $\chi$. Volumetric growth of the 3D cell collective is significantly enhanced at $\chi=1$ as compared to $\chi=0$.}%\datta{The images combined in PPT have different scales. They all need to be the same size. When generating images in MATLAB the figure size can be controlled, I think}}
\label{fig:rg2b}
\end{figure}

The cell softening probability clearly determines the cell dynamics as evident from the MSD dependence on $\chi$ (discussed above). We next sought to 
evaluate whether cell softening impacts the volumetric growth of tumor cell collectives. Finding the biophysical underpinnings of tumor growth is of much interest. This is an important problem because accurate tumor growth modeling can be crucial in evaluating patient screening strategies~\cite{talkington2015estimating}, establishing radiation treatment protocols~\cite{castorina2007growth} as well as assist treatment decisions~\cite{comen2012translating}.  
To answer this question, we quantified the 3D spatial spread of the cell collective using radius of gyration squared,  
\begin{equation}
      R_g^2(t)=\left\langle  \frac{1}{N} \Sigma_{i=1}^{N} [{\bf r}_{i}(t)-{\bf R}_{CM}(t)]^2 \right\rangle.
      \label{eq:rg2}
\end{equation} 
The bracket $\langle ... \rangle$ denotes ensemble average over 12 different simulation runs at each value of $\chi$ for different initial conditions (see Appendix~\ref{sec:appendixA}). The average squared distance of all the cells from the center of mass gives a sense of the size of the 3D cell collective. 
%in Fig.~\ref{fig:rg2}. 
Small $R_g^2$ values indicate cell positions that are localized 
in close proximity to the center of mass. In contrast, cells spatially  
distributed farther away from the center of mass leads to 
significantly larger $R_g^2$ 
values~\cite{saxton1993lateral,roy2017haptotaxis}. As a result, $R_g^2(t)$ as a function of time is a readout of the 3D cell collective volumetric growth. 
The time varying $R_g^2$ in 
Fig.~\ref{fig:rg2a} shows slow change at $t<\sim 3$ days followed by faster growth at $t>4$  
days. The $R_g^2$ values are indistinguishable between 
$\chi$ values at time below $4$ days as compared to 
later times when $R_g^2$ is significantly larger for  
$\chi=1$. Time dependent  scaling of $R_g^2$ based on a  fit to power law $R_g^2(t)\sim t^{\beta}$ reveals  
important features of cell spatial distribution 
dynamics in 3D~\cite{parry2014bacterial,gonzalez2008understanding} (see Inset of Fig.~\ref{fig:rg2a}). When $\beta=1$, 
cells exhibit diffusive random spread compared to when 
cells undergo directed  
spreading at $\beta>1$. By contrast restricted cell 
spreading leads to sublinear rise in $R_g^2$ with 
$\beta<1$. 
The maximum (max) $R_g^2$ values (marked as stars in Inset of Fig.~\ref{fig:rg2a}) show a clear linear 
trend with $\chi$ (see Fig.~\ref{fig:rg2b}a). This implies that enhanced mechanical heterogeneity leads to significantly more spread out morphology of the 
3D cell collective. The median value of max $R_g^2$ at $\chi=0.1$ is $\sim 5800~\mu m^2$ as compared to 
$\sim 7100~\mu m^2$ at $\chi=1$ as shown in Fig.~\ref{fig:rg2b}a. 

Our results therefore indicate that heightened spatial mechanical heterogeneity leads to enhanced volumetric growth of the 3D cell collective with 
time dependent spatial expansion of the cell collective being restricted when individual cells are stiffer. In contrast, cell softening favored faster expansion of the cell collective into the surrounding viscous medium  with a median value of $\beta\sim 1.2$ 
(Fig.~\ref{fig:rg2b}b). Our results provide evidence 
into how spatial mechanical heterogeneity determines the spatial spread of 3D cell collectives. 
Hence, spatial mechanical heterogeneity 
consisting of a 
stiffer core cells and softer peripheral cells aid  in 
more efficient volumetric growth of cell collectives. 

 \section{\label{sec:secVI}Conclusion}
Understanding how individual cell level mechanical changes impact cell dynamics and tumor growth is critical to understanding cancer progression. In this respect, we studied how time varying individual cell stiffness drives 
spatial %\datta{in what sense is it topological - will this predicted behavior hold even for cylindrical cancer cells} 
mechanical heterogeneity in multicellular collectives by incorporating stiffening of cells 
immediately prior to division and softening post division into our minimal 3D  tumor growth model. The probability for cells to  soften after division is implemented in our model through the parameter $\chi$ which tunes actin   
rearrangement from the cell cortex into the cytoplasm.  Our simulations show that cell division associated softening drives the emergence of spatial mechanical heterogeneity between the core and periphery of multicellular spheroids. The resulting spatial stiffness pattern consisting of a core made up of stiffer cells and peripheral softer cells enhances the 3D  collective cell dynamics and volumetric growth of multicellular spheroids.  Broadly, our computational results are consistent with experimental observations of spatial mechanical heterogeneity in 3D tumor organoids~\cite{han2020cell}, and the heightened ability of softer tumor cells to metastasize~\cite{swaminathan2011mechanical,panzetta2017mechanical}. As polymerization and depolymerization of the actomyosin network in the cell cortex leads to time varying stiffening and softening of the cell, we show that such temporal stiffness variation at the single cell level is essential in the emergence of mechanical heterogeneity. In addition to the increased space that cells explore in 3D cell collectives due to periodic stiffening and softening, our study shows that increased spatial mechanical heterogeneity is correlated with enhanced 3D spheroid growth. 
Our results therefore have important implications into understanding how time variations in single cell mechanical 
properties determine the spatial organization and dynamics at the cell collective scale. 
%\datta{If periodic cell stiffening and softenting is a universal cancer process phenomenon, then this simulation points to a very universal relation, irrespective of the cancer type. Somewhere in the discussion we may want to say that, if the assumptions are true.} 

\begin{acknowledgments}
A.M.K acknowledge funding from startup grants. A.M.K and T.~D. acknowledges funding support from Augusta University CURS Summer Scholars Program. The authors acknowledge the support of the Augusta University High Performance Computing Services (AUHPCS) for providing computational resources contributing to the results presented in this publication. We thank Sumit Sinha, Xin Li and Dave Thirumalai for valuable comments on the manuscript. 
\end{acknowledgments}

\FloatBarrier  
\appendix
\section{Initial Conditions}\label{sec:appendixA}
%\datta{Shouldn't this be part of the main text?} I think this is fine as part of the appendix
We initiated the simulations by placing 100 cells whose $x$, $y$, $z$ coordinates are chosen from a normal distribution with zero mean and standard deviation $40~\mu m$. In the present study, all the individual cell parameters are fixed except single cell stiffness $E_i$ which is varied within a physiological cell stiffness range. The simulated dense 3D cell aggregate was evolved for $650,000$~sec or $12\tau$. At each $\chi$ value, 12 different simulation runs allow for random initial positions of cells. Hence, our reported results account for varying initial conditions. Relevant simulation parameters are shown in Table~1. The time-dependent coordinates of particles were recorded to calculate the dynamical observables relevant to this study. 

\section{Simulation movies}\label{sec:appendixB}
Movies generated from the simulated 3D cell collective are shown. The total duration of the movie is $650,000$~sec or $\approx12 \tau$. The time interval between consecutive frames is $1000~\text{sec}$.\\
{\bf Movie 1:} 3D cell collective simulated at $\chi=1$. Color bar indicates the stiffness of cells with dark blue indicating stiffer cells at $3KPa$. Softer cells are show in yellow color at $0.5KPa$. The observation frame is rotated to allow for a full 3D view of the tumor cell collective. The box is for 3D visualization purposes only. %\datta{Create a tiny url and host from your webpage?}
\href{https://augustauniversity.box.com/s/zyse7otymd2uu2zt451vgx6g3x28abwa}{(Link)} \\
{\bf Movie 2:} Cross-section view of a 3D cell collective simulated at $\chi=1$. Color bar indicates the stiffness of cells with dark blue indicating stiffer cells at $3KPa$. Softer cells are shown in yellow color at $0.5KPa$. A view of a fixed 2D plane cutting through the 3D cell collective is shown. 
\href{https://augustauniversity.box.com/s/xxa0eezs2rae2yj7shlp7kyk7dvx19m0}{(Link)}

%\datta{include movie as supplemental material link, with the text of movie description provided with supplementary?}\\
%I have typically included it in the appendix 
% {\bf Movie 2:} Particle aggregates simulated for $f^{ad}=2\times 10^{-4}~\mathrm{\mu N/\mu m^{2}}$. \href{https://drive.google.com/file/d/1BLOlwQDbUBJCAm84EjtknaaQKXm3z9I8/view?usp=sharing}{(Link)}\\
% {\bf Movie 3:} Particle aggregates simulated for $f^{ad}=3\times 10^{-4}~\mathrm{\mu N/\mu m^{2}}$. \href{https://drive.google.com/file/d/1QUSpyiHtvSvkhAgM6gsfylXnVdXGHg5h/view?usp=sharing}{(Link)}. \\

\bibliography{ms}% Produces the bibliography via BibTeX.

\end{document}